\documentclass[11pt,epsf]{article}
\newlength{\vshift}
\input epsf.tex
\newlength{\hshift}
\usepackage{amssymb}
\usepackage{amsmath}
\usepackage{amsthm}
\usepackage{amsopn}
\def\nn{\nonumber }
\def\la{\lambda}
\def\La{\Lambda}

\def\th{\theta}
\def\ka{\kappa}
\def\de{\delta}
\def\be{\beta}
\def\ga{\gamma}
\def\al{\alpha}
\def\si{\sigma}
\def\ep{\epsilon}
\def\ds{\stackrel{\star}{,}}
\def\x{\hat x}
\def\p{\partial}

\def\lb{\lbrack}
\def\rb{\rbrack}
\def\Lo{\overline{L}^i}
\def\Loo{{{\overline{L}}^1}^i}
\def\Lot{{{\overline{L}}^2}^i}
\def\beq{\begin{equation}}
\def\eeq{\end{equation}}
\def\bea{\begin{eqnarray}}
\def\eea{\end{eqnarray}}
\begin{document}

\vspace{4em}
\begin{center}

{\Large{\bf Lorentz Conserving Noncommutative Standard Model}}

\vskip 3em

{{\bf  M. M. Ettefaghi \footnote{e-mail: ettefaghi@ph.iut.ac.ir }
, M. Haghighat \footnote{e-mail: mansour@cc.iut.ac.ir }}}

\vskip 1em

Department of Physics Isfahan University of technology, Isfahan,
Iran
 \end{center}
\begin{abstract}
We consider Lorentz conserving noncommutative field theory to
construct the Lorentz conserving noncommutative standard model based
on the  gauge group $SU(3)\times SU(2)\times U(1)$.  We obtain the
enveloping algebra-valued of Higgs field up to the second order of
the noncommutativity parameter $\th_{\mu\nu}$.  We derive the action
at the leading order and find new vertices which are absent in the
ordinary Standard Model as well as the minimal noncommutative
standard model. We briefly study the phenomenological aspects of the
model.
\end{abstract}
\section{Introduction}
\label{A} In recent years, many authors have considered
noncommutative (NC) field theories and their phenomenological
aspects \cite{nc}. A strong motivation for investigating these field
theories is their appearance in a definite limit of string theory
\cite{sw}. On the other hand the standard model of electro-weak and
strong interactions has met the challenge of many high precision
experiments. In the high energy limit the noncommutativity effects
seem to be significant and therefore the new interactions
 in the noncommutative space and time can be
potentially important to particle physics and cosmology. For example
in the minimal extension of the standard model in the noncommutative
space, in contrast with the conventional theory, there is
neutrino-photon vertex which leads to neutrino-photon interaction at
the tree level \cite{hez}. In the canonical noncommutative
space-time, the coordinates are operators and satisfy the following
commutator relation
 \beq \label{ncalgebra}
  \lb \x^\mu,\x^\nu \rb=i\th^{\mu\nu},
 \eeq

 where $\th^{\mu\nu}=-\th^{\nu\mu}$ is real and constant
Lorentz tensor. As $\th^{\mu\nu}$ is constant, there is, obviously,
a preferred direction in a given particle Lorentz frame which leads
to the Lorentz symmetry violation. On the other hand, experimental
inspections for Lorentz violation, including clock comparison tests,
polarization measurements on the light from distant galaxies,
analyses of the radiation emitted by energetic astrophysical
sources, studies of matter-antimatter asymmetries for trapped
charged particles and bound state systems \cite{cc} and so on, have
thus far failed to produce any positive results.  These experiments
strictly bound the Lorentz-violating parameters, therefore, in the
lower energy limit, the Lorentz symmetry is an almost exact symmetry
of the nature. However, it is natural to explore the noncommutative
field theories that are Lorentz invariant from the beginning. In
this class of NC theories, the parameter of noncommutativity is not
a constant but an operator which transforms as a Lorentz tensor. Of
course in this way one needs to generalize the star product and
operator trace for functions of both $x^\mu$ and $\th^{\mu\nu}$,
appropriately. However, in both cases experiment should confirm the
theories. Using enveloping algebra-valued method, introduced in
\cite{ncg,lm}, Carlson, Carone and Zobin (CCZ) have constructed
Lorentz conserving noncommutative quantum electrodynamics based on a
contracted Snyder algebra \cite{ccz}. Afterward, the miscellaneous
aspects of the theory has been considered by others
\cite{he,kn,cn,ck}. In this paper we introduce Lorentz conserving
noncommutative standard model (LCNCSM) using the CCZ approach and
consider differences between the LCNCSM and the Lorentz violating
noncommutative standard model. To construct the noncommutative field
theory, according to the Weyl- Moyal correspondence, ordinary
function can be used instead of the corresponding noncommutative one
by replacing the ordinary product with the star product as follows
\begin{equation}
f*g(x)=f(x)exp(i/2\overleftarrow{\partial_{\mu}}\th^{\mu\nu}
\overrightarrow{\partial_{\nu}})g(x).
\end{equation}
Using this correspondence, however, there are two approaches to
construct the gauge theories in the noncommutative space. In the
first one the gauge group is restricted to $U(n)$ and the symmetry
group of the standard model is achieved by reduction of $U(3)\times
U(2)\times U(1)$ to $SU(3)\times SU(2)\times U(1)$ by an appropriate
symmetry breaking \cite{nogo}. Nevertheless, for $U(1)$ the charge
of particles are allowed to be $\pm1$ and $0$ \cite{haya,nogo}. In
the second approach, the noncommutative gauge theory can be
constructed for every charges and for $SU(n)$ gauge group via
Seiberg-Witten map\cite{ncg,lm}. Meanwhile to follow the second
approach to construct LCNCSM one needs the Seiberg-Witten map of all
 fields in the standard model up to the second order of
$\th^{\mu\nu}$.  Fortunately the map for all of the fields except
the Higgs field to this order has already been obtained \cite{lm}.

In Sec. II we briefly review Lorentz conserving noncommutative field
theory. In Sec. III  we study enveloping algebra and the
Seiberg-Witten map for the fields of the standard model and obtain
the corresponding expression for the Higgs field up to the second
order of $\th^{\mu\nu}$ which is not calculated elsewhere. In Sec.
IV we construct the LCNCSM and discuss possible vertices in this
model. Finally, we discuss the phenomenological aspects of this
model in Sec.V, and give the concluding remarks in Sec.VI.
\section{Lorentz Conserving Noncommutative Field Theory}
In 1947 Snyder considered the Lorentz symmetry in discrete
space-time to avoid UV divergence \cite{snyder}. For this purpose,
he assumed that the space-time coordinates are non-commutative
operators which led to a Lorentz-invariant discrete space-time. CCZ
by contracting the proposed algebra found the following algebra \bea
 &&[\widehat{x}^{\mu},\widehat{x}^{\nu}]=i\widehat{\th}^{\mu\nu},
 \nn\\
 &&[\widehat{\th}^{\mu\nu},\widehat{x}^\la]=0, \nn\\
 &&[\widehat{\th}^{\mu\nu},\widehat{\th}^{\al\be}]=0,
\eea which is similar to the canonical noncommutative algebra but
$\widehat{\th}^{\mu\nu}$ is  an antisymmetric operator that is not
constant but transforms as a Lorentz tensor. The action for Lorentz
conserving field theories on noncommutative spaces are then obtained
using the Weyl-Moyal correspondence.  In fact, in order to find the
noncommutative action, the usual product of fields should be
replaced by the star product:
 \beq
f*g(x,\th)=f(x,\th)\exp(i/2\overleftarrow{\partial_{\mu}}\th^{\mu\nu}
\overrightarrow{\partial_{\nu}})g(x,\th).\label{star}
 \eeq
 It should be noted that here the mapping to c-number coordinates
involve ${\th}^{\mu\nu}$ as a c-number due to the presence of the
operator $\widehat{\th}^{\mu\nu}$ in the Lorentz-conserving case. In
this formulation, the operator trace that is a map from operator
space to numbers, is defined as
 \beq
Tr\hat{f}=\int d^{4}xd^{6}\th W(\th)f(x,\th),
 \eeq
where $W(\th)$ is a Lorentz invariant weight function with the
normalization $\int d^{6}\th W(\th)=1$,
 and is assumed to be a positive and even function of $\th_{\mu\nu}$.  Therefore
 \beq
\int{ d^6\th W(\th)\th^{\mu\nu}}=0.
 \eeq
Also for every even Lorentz invariant weighting function $W(\th)$
one has
 \beq
\int d^6\th W(\th)\th^{\mu\nu}\th^{\ka\la}=
\frac{\langle\th^2\rangle}{12}(g^{\mu\ka}g^{\nu\la}-g^{\mu\la}g^{\nu\ka}),
 \eeq
where
 \beq
\langle\th^2\rangle=\int d^6\th
W(\th)\th^{\mu\nu}\th_{\mu\nu}.
 \eeq
 Furthermore, the weight function is assumed to fall sufficiently
fast so that all integrals are well defined.  In fact $W(\th)$
suppresses cross-section for center-of-mass energy beyond the value
of noncommutative scale, therefore working with truncated power
series expansion of functions in $\th_{\mu\nu}$ is permitted.  Now
the properties of $W(\th)$ and the definition of the operator trace
allows one to obtain the Lagrangian for the Lorentz conserving
noncommutative field theory
 \beq
 {\cal{L}}(x)=\int{d^6\th W(\th){\cal{L}}(\phi,\partial\phi)_*},
 \eeq
 in which ${\cal{L}}(\phi,\partial\phi)_*$ depends on both $x$ and $\th_{\mu\nu}$, and its subscript
indicates the *-product defined in Eq.(\ref{star}) .
\section{$\theta^{\mu\nu}$-Expanded Fields up to the Second Order}
 A non-Abelian gauge theory, based on a Lie group, for example $SU(n)$, in the
 noncommutative space can
not be constructed in the same way as the commutative one. In fact
the main difference is that for every two gauge parameters , $\La$
and $\La'$, one has
 \beq
[\La\ds\La']=\frac{1}{2}\lb
T^a,T^b\rb\{\La_{1,a}(x)\ds\La_{2,b}(x)\}+\frac{1}{2}\{
T^a,T^b\}\lb\La_{1,a}(x)\ds\La_{2,b}(x)\rb,
 \eeq
 where $\La={{\La}_a}(x)T^a$.  Obviously, the anti-commutators of
 $T^a$'s do not close the Lie algebra of a non-Abelian gauge theory
 except for $U(N)$ and they reproduce all the higher powers of the generators.  Meanwhile the
enveloping algebra consists of all ordered tensor powers of the
generators $T^a$ and seems to be a proper choice for such a gauge
theory \cite{ncg,lm}.  However, the enveloping algebra is
infinite-dimensional and as a consequence the enveloping
algebra-valued noncommutative gauge parameter and fields would have
an infinite number of degrees of freedom which can be considered as
follows
 \bea \label{ela}
&{\widehat{\La}_\al}&\!\!\!\!\!=\al+\hbar \La_\al^1 +\hbar^2\La_\al^2+\dots, \\
\label{epsi}
&{\widehat{\Psi}}&\!\!\!\!\!=\Psi^0+\hbar \Psi^1 +\hbar^2\Psi^2+\dots,\\
\label{eA}&{\widehat{A}_\mu}&\!\!\!\!\!=A^0_\mu+\hbar
A^1_\mu+\hbar^2A^2_\mu+\dots,\\ \label{ephi}
&{\widehat{\Phi}}&\!\!\!\!\!=\Phi^0+\hbar\Phi^1+\hbar^2\Phi^2+\dots,
 \eea
where $\alpha$ is the ordinary gauge parameter and $ \Psi^0$, $
A^0_\mu$ and $ \Phi^0$ are respectively the commutative fermion
fields, gauge fields and Higgs fields, and the superscript $i$
stands for their order in the expansion. The infinite number of
degrees of freedom can be restricted demanding that the enveloping
algebra valued quantities ( such as gauge and matter fields and so
on) depend on the algebra valued ones and their space-time
derivatives only. This requirement, based on existence of the
Seiberg-Witten map to all orders, reduces the number of degrees of
freedom of the NC gauge theory to the same one of the gauge theory
of the commutative space. In fact Seiberg and Witten have shown that
there is an equivalence between ordinary and noncommutative gauge
fields to any finite order in $\th_{\mu\nu}$ which is realized by a
map in a way that preserves the gauge equivalence relation.  In the
other words if $\hat{A}_\mu$ and $\hat{A'}_\mu$ are equivalent gauge
fields in noncommutative space-time, the corresponding ordinary
gauge fields $A_\mu$ and $A'_\mu$ should be equivalent too. This
means
 \beq
\de\hat{A}_\mu=\hat{\de}\hat{A}_\mu \,\,\,\, , \,\,\,\,
\de\hat{\Psi}=\hat{\de}\hat{\Psi}.
 \eeq
In the ordinary space the commutator of two infinitesimal gauge
transformations are closed. Therefore, it is necessary to consider
the following consistency condition for the noncommutative gauge
parameter
 \beq \label{consistency}
i\de_\al{\widehat{\La}}_\be-i\de_\be{\widehat{\La}}_\al+[{\widehat{\La}}_\al\ds{\widehat{\La}}_\be]
=i{\widehat{\La}}_{-i[\al,\be]}.
 \eeq
 By substituting the expansion of ${\widehat{\La}}_\al$ from (\ref{ela}) in (\ref{consistency}), one up to the first order of $\th_{\mu\nu}$ finds
 \beq
i(\de_\al{\La^1}_\be-\de_\be{\La^1}_\al)+[\al,{\La^1}_\be]+[{\La^1}_\al,\be]-i{\La^1}_{-i[\al,\be]}=
\frac{i}{2}\th^{\mu\nu}\{\p_\mu\al,\p_\nu\be\},
 \eeq
 and after a little algebra up to the second order of $\th_{\mu\nu}$ one
 has
 \bea
&i&\!\!\!\!\!\!\!(\de_\al{\La^2}_\be-\de_\be{\La^2}_\al)+[\al,{\La^2}_\be]+
[{\La^2}_\al,\be]-i{\La^2}_{-i[\al,\be]}=
\frac{1}{8}\th^{\mu\nu}\th^{\ka\la}[\p_\mu\p_\ka\al,\p_\nu\p_\la\be]\nn\\&-&
 \!\!\!\!\!\![{\La^1}_\la,{\La^1}_\be]
-\frac{i}{2}\th^{\mu\nu}(\{\p_\mu{\La^1}_\al,\p_\nu\be\}-\{\p_\mu{\La^1}_\be,\p_\nu\al\}),
 \eea
  which in terms of ${\La^1}_\al$ and ${\La^2}_\al$  are inhomogeneous linear
equations with the following solutions, respectively
\cite{ncg,lm}:
 \beq
{\La}^1_\al=-\frac{1}{4}\th^{\mu\nu}\{A^0_\mu,\p_\nu\al\},\label{la1}
 \eeq
 and
 \bea
\La^2_\al=\!\!\!\!\!&{\frac{1}{32}}&\!\!\!\!\!\th^{\mu\nu}\th^{\ka\la}(\{A^0_\mu,\{\p_\nu
A^0_\ka,\p_\la\al\}\}
+\{A^0_\mu,\{A^0_\ka,\p_\nu\p_\la\al\}\}+\{\{A^0_\mu,\p_\nu
A^0_\ka\},\p_\la\al\}\nn\\
&-&\!\!\!\!\!\{\{F^0_{\mu\ka},A^0_\nu\},\p_\la\al\}-2i[\p_\mu
A^0_\ka,\p_\nu\p_\la\al])\label{la2}.
 \eea
 In the second step the noncommutative field $\widehat{\Psi}$ can be determined
 by replacing the expansion of $\widehat{\La}_\al$ and $\widehat{\Psi}$ which
 is given, respectively, in (\ref{ela}) and (\ref{epsi}), in the gauge
transformation $\de\widehat{\Psi}=i\widehat{\La}_\al*\widehat{\Psi}$
\cite{ncg,lm}.  It can be easily shown that up to the first order of
$\th_{\mu\nu}$ one has
 \beq
\de_\al\Psi^1-i\al\Psi^1=i\La^1_\al\Psi^0-\frac{1}{2}\th^{\mu\nu}\p_\mu\al\p_\nu\Psi^0,
 \eeq
and up to the second order of $\th_{\mu\nu}$, it results in
 \bea
\de_\al\Psi^2-i\al\Psi^2=\!\!\!\!\!&i&\!\!\!\!\!\!\!\La^2_\al\Psi^0+
i\La^1_\al\Psi^1-\frac{1}{2}\th^{\mu\nu}\p_\mu\La^1_\al\p_\nu\Psi^0
-\frac{1}{2}\th^{\mu\nu}\p_\mu\al\p_\nu\Psi^1\nn\\&-
&\!\!\!\!\!\frac{i}{8}\th^{\mu\nu}\th^{\ka\la}\p_\mu\p_\ka\al\p_\nu\p_\la\Psi^0.
 \eea
 The first equation can be solved by
 \beq
\Psi^1=-\frac{1}{2}\th^{\mu\nu}A^0_\mu\p_\nu\Psi^0+\frac{i}{4}\th^{\mu\nu}A^0_\mu
A^0_\nu\Psi^0,\label{psi1}
 \eeq
  and the solution of the second equation can be obtained as follows
 \bea
\label{psi2} \Psi^2=-\frac{i}{8}\th^{\mu\nu}\th^{\ka\la}(\p_\ka
A^0_\mu\p_\nu\p_\la\Psi^0+iA^0_\ka
A^0_\mu\p_\nu\p_\la\Psi^0-i\p_\ka A^0_\mu
A^0_\nu\p_\la\Psi^0+iF^0_{\ka\mu}A^0_\nu\p_\la\Psi^0\nn\\
-iA^0_\nu\p_\ka A^0_\mu\p_\la\Psi^0+ 2iA^0_\nu
F^0_{\ka\mu}\p_\la\Psi^0+2A^0_\mu A^0_\ka
A^0_\nu\p_\la\Psi^0-A^0_\mu A^0_\nu A^0_\ka\p_\la\Psi^0\nn\\
-\frac{i}{4}(2\p_\ka A^0_\mu\p_\la A^0_\nu\Psi^0-2i\p_\ka A^0_\mu
A^0_\la A^0_\nu\Psi^0+2iA^0_\nu A^0_\la\p_\ka A^0_\mu\Psi^0
+i[[\p_\ka A^0_\mu,A^0_\nu],A^0_\la]\Psi^0\nn\\+4iA^0_\nu
F^0_{\ka\mu}A^0_\la\Psi^0-A^0_\ka A^0_\la A^0_\mu A^0_\nu\Psi^0
+2A^0_\ka A^0_\mu A^0_\nu A^0_\la\Psi^0))\label{psi22}.
 \eea
 The enveloping algebra gauge potential can be found by inserting $\widehat{A}$ and
$\widehat{\La}_\al$ from (\ref{eA}) and (\ref{ela}), respectively,
in the transformation $\de
\widehat{A}_\mu=\p_\mu\widehat{\La}_\al-i[\widehat{A}_\mu\ds\widehat{\La}_\al]$
and retaining the coefficients up to the first order of
$\th_{\mu\nu}$ leads to
 \beq
\de_\al
A^1_\si-i[\al,A^1_\si]=\p_\si\La^1_\al-i[A^0_\si,\La^1_\al]+\frac{1}{2}\th^{\mu\nu}\{\p_\mu
A^0_\si,\p_\nu\al\},
 \eeq
 which has a solution as follows
 \beq A^1_\si=-\frac{1}{4}\th^{\mu\nu}(\{A^0_\mu,\p_\nu
A^0_\si\}-\{F^0_{\mu\si},A^0_\nu\})\label{a1}.
 \eeq
 For the next order of $\th_{\mu\nu}$ after some manipulation one has
  \bea
\de_\al
A^2_\si-i[\al,A^2_\si]=\p_\si\La^2_\al-i[A^0_\si,\La^2_\al]-i[A^1_\si,\La^1_\al]+\frac{1}{2}\th^{\mu\nu}\{\p_\mu
A^1_\si,\p_\nu\al\}\nn\\
+\frac{1}{2}\th^{\mu\nu}\{\p_\mu
A^0_\si,\p_\nu\La^1_\al\}+\frac{i}{8}\th^{\mu\nu}\th^{\ka\la}[\p_\ka\p_\mu
A^0_\si,\p_\la\p_\nu\al],
 \eea
 with the solution given in the following equation.  There are differences in signs of
 some terms of given equation in comparison with the corresponding
 one given in reference \cite{lm}. The misprinting in the signs can be easily
 verified by considering the reduction of the equation to the
 Abelian case.
 \bea
A_{\si}^{2}=&&\!\!\!\!\!\!\!\!\!\!\frac{1}{32}\theta^{\mu\nu}\theta^{\ka\la}\Big(\{\{A^0_\ka,\partial_\la
A^0_\mu\},\partial_\nu A^0_\si\}-\{\{F^0_{\ka\mu},
A^0_\la\},\partial_\nu A^0_\si\}-2i[\partial_\ka
A^0_\mu,\partial_\la\partial_\nu A^0_\si]\nonumber\\
&&\quad -\{A^0_\mu, \{\partial_\nu F^0_{\ka \si},
A^0_\la\}\}-\{A^0_\mu, \{ F^0_{\ka \si}, \partial_\nu
A^0_\la\}\}  +
\{A^0_\mu, \{\partial_\nu A^0_\ka, \partial_\la A^0_\si\}\}\nonumber\\
&&\quad  + \{A^0_\mu, \{ A^0_\ka, \partial_\nu\partial_\la
A^0_\si\}\} - \{\{A^0_\ka,\partial_\la F^0_{\mu\si}\}, A^0_\nu\}
+\{\{\mathcal{D}_\ka^0 F^0_{\mu\si}, A^0_\la\}, A^0_\nu\}\label{a2} \\
&&\!\!\!\!\!\!\!\!\!\!\!\!\!\!+2\{\{F^0_{\mu\ka},F^0_{\si\la}\},
A^0_\nu\}+2i[\partial_\ka F^0_{\mu\si},\partial_\la
A^0_\nu]-\{F^0_{\mu\si},\{A^0_\ka,
\partial_\la A^0_\nu\}\}+\{F^0_{\mu\si},\{F^0_{\ka
  \nu},A^0_\la\}\}\Big), \nonumber
 \eea
 where $F^0_{\mu\nu}=\p_\mu A^0_\nu-\p_\nu A^0_\mu
-i[A^0_\mu,A^0_\nu]$ is the Lie algebra field strength.  Finally,
to construct LCNCSM we need to determine the hybrid
Seiberg-Witten map for Higgs fields. The gauge transformation for
Higgs field which transforms on the left and on the right under
two arbitrary gauge groups with corresponding gauge potentials is
 \bea
\de\widehat{\Phi}=i\widehat{\La}_\al*\widehat{\Phi}-i\widehat{\Phi}*\widehat{\La}'_\ga,
 \eea
  in which  $\widehat{\Phi}$, $\widehat{\La}_\al$ and $\widehat{\La}'_\ga$ are
  defined by the Eqs.(\ref{ela}) and (\ref{ephi}).  Therefore, the noncommutative
   Higgs transformation at the first order of $\th_{\mu\nu}$  reduces to
   \beq
 \de\Phi^1-i\al\Phi^1+i\Phi^1\ga=-\frac{1}{2}\th^{\mu\nu}\p_\mu\Phi^0\p_\nu\Phi^0+\frac{1}{2}\th^{\mu\nu}\p_\mu\Phi^0\p_\nu\ga
 +i\La^1_\al\Phi^0-i\Phi^0\La^{1'} _\ga,
  \eeq
   with the solution
 \beq
 \Phi^1=\frac{1}{2}\th^{\mu\nu}A^0_\nu(\p_\mu\Phi^0-\frac{i}{2}(A^0_\mu\Phi^0-\Phi^0A^{0'}_\mu))
 +\frac{1}{2}\th^{\mu\nu}(\p_\mu\Phi^0-\frac{i}{2}
 (A^0_\mu\Phi^0-\Phi^0A^{0'}_\mu))A^{0'}_\nu,\label{phi1}
 \eeq
 and at the next order leads to
    \bea
 \de\Phi^2-i\al\Phi^2+i\Phi^2\ga=\!\!\!\!\!&{\frac{i}{8}}&\!\!\!\!\!\th^{\mu\nu}\th^{\ka\la}\p_\mu\p_\ka\Phi^0\p_\nu\p_\la\ga
 -\frac{i}{8}\th^{\mu\nu}\th^{\ka\la}\p_\mu\p_\ka\al\p_\nu\p_\la\Phi^0
 -\frac{1}{2}\th^{\mu\nu}\p_\mu\La^1_\al\p_\nu\Phi^0\nn\\&-&\!\!\!\!\!\frac{1}{2}\th^{\mu\nu}\p_\mu\al\p_\nu\Phi^1
 +\frac{1}{2}\th^{\mu\nu}\p_\mu\Phi^0\p_\nu\La^{1'}_\ga+\frac{1}{2}\th^{\mu\nu}\p_\mu\Phi^1\p_\nu\ga+i\La^1_\al\Phi^1
 \nn\\&+&\!\!\!\!\!i\La^2_\al\Phi^0-i\Phi^0\La^{2'}_\ga-\Phi^1\La^{1'}_\ga,
  \eea
   which can be solved up to this order as
 \beq
\Phi^2=\Phi^2_1+\Phi^2_2+\Phi^2_3 \label{phi2},
 \eeq
  where $\Phi^2_1=\Psi^2$ and is given in Eq.(\ref{psi2}),
 \bea
\Phi^2_2=\th^{\mu\nu}\th^{\ka\la}(&&\!\!\!\!\!\!\!\!\!\!\!-\frac{3}{32}\Phi^0{A^0}_\la'{A^0}_\ka'{A^0}_\nu'{A^0}_\mu'
+\frac{1}{8}\Phi^0{A^0}_\la'{A^0}_\nu'{A^0}_\ka'{A^0}_\mu'
-\frac{1}{16}\Phi^0{A^0}_\la'{A^0}_\nu'{A^0}_\mu'{A^0}_\ka'\nn\\&&\!\!\!\!\!\!\!\!\!\!\!-\frac{5i}{32}\Phi^0{A^0}_\la'\p_\mu
{A^0}_\ka'{A^0}_\nu'-\frac{i}{32}\Phi^0\p_\mu
{A^0}_\la'{A^0}_\ka'{A^0}_\nu'-\frac{i}{8}\p_\mu\Phi^0{A^0}_\la'{A^0}_\ka'{A^0}_\nu'
\nn\\&&\!\!\!\!\!\!\!\!\!\!\!-\frac{i}{16}\Phi^0{A^0}_\la'{A^0}_\nu'\p_\mu
{A^0}_\ka'-\frac{i}{16}\Phi^0\p_\mu
{A^0}_\la'{A^0}_\nu'{A^0}_\ka'+\frac{i}{8}\p_\mu\Phi^0{A^0}_\la'{A^0}_\nu'{A^0}_\ka'
\nn\\&&\!\!\!\!\!\!\!\!\!\!\!+\frac{i}{32}\Phi^0{A^0}_\la'{A^0}_\nu'\p_\ka
{A^0}_\mu'-\frac{3i}{32}\Phi^0{A^0}_\la'\p_\ka
{A^0}_\nu'{A^0}_\mu'-\frac{i}{8}\p_\ka\Phi^0{A^0}_\la'{A^0}_\nu'{A^0}_\mu'
\nn\\&&\!\!\!\!\!\!\!\!\!\!\!-\frac{1}{16}\Phi^0\p_\nu
{A^0}_\la'\p_\mu {A^0}_\ka'-\frac{1}{8}\p_\nu\Phi^0\p_\mu
{A^0}_\la'{A^0}_\ka'+\frac{1}{8}\p_\mu\p_\ka\Phi^0{A^0}_\la'{A^0}_\nu'\nn\\&&\!\!\!\!\!\!\!\!\!\!\!\!+\frac{1}{4}\p_\ka\Phi^0\p_\mu
{A^0}_\la'{A^0}_\nu'+\frac{1}{8}\p_\mu\Phi^0{A^0}_\la'\p_\ka'{A^0}_\nu'-\frac{i}{8}\p_\nu\p_\ka\Phi^0\p_\mu
{A^0}_\la'),
 \eea
and $\Phi^2_3$ is
\bea
\Phi^2_3=\th^{\mu\nu}\th^{\ka\la}(&&\!\!\!\!\!\!\!\!\!\!\!\!\!\!\!\!-\frac{i}{4}{A^0}_\nu\p_\mu
{A^0}_\ka\Phi^0{A^0}_\la'+\frac{i}{8}{A^0}_\mu {A^0}_\nu
\p_\ka\Phi^0{A^0}_\la'-\frac{i}{8}{A^0}_\ka {A^0}_\nu\Phi^0\p_\mu
{A^0}_\la'\nn\\&&\!\!\!\!\!\!\!\!\!\!\!\!\!\!\!\!-\frac{i}{4}{A^0}_\ka\Phi^0\p_\mu
{A^0}_\la'{A^0}_\nu'+\frac{i}{8}{A^0}_\nu\p_\mu\Phi^0{A^0}_\ka'{A^0}_\la'-\frac{i}{8}\p_\mu
{A^0}_\ka\Phi^0{A^0}_\nu'{A^0}_\la'\nn\\&&\!\!\!\!\!\!\!\!\!\!\!\!\!\!\!\!
+\frac{1}{8}{A^0}_\mu{A^0}_\nu{A^0}_\ka\Phi^0{A^0}_\la'
-\frac{1}{8}{A^0}_\nu\Phi^0{A^0}_\mu'{A^0}_\ka'{A^0}_\la'+\frac{1}{4}\p_\mu{A^0}_\ka\Phi^0\p_\nu
{A^0}_\la'\nn\\&&\!\!\!\!\!\!\!\!\!\!\!\!\!\!\!\!
+\frac{1}{4}{A^0}_\nu\p_\mu\p_\al\Phi^0{A^0}_\be'+\frac{1}{8}\p_\ka
{A^0}_\nu\Phi^0\p_\mu {A^0}_\la'-\frac{1}{8}{A^0}_\mu
{A^0}_\ka\Phi^0{A^0}_\nu'{A^0}_\la'\nn\\&&\!\!\!\!\!\!\!\!\!\!\!\!\!\!\!\!-\frac{1}{16}{A^0}_\mu
{A^0}_\nu\Phi^0{A^0}_\ka'{A^0}_\la'+\frac{i}{8}{F^0}_{\ka\mu}{A^0}_\nu\Phi^0{A^0}_\la'
+\frac{i}{8}{A^0}_\ka\Phi^0{A^0}_\nu'{F^0}_{\la\mu}'
\nn\\&&\!\!\!\!\!\!\!\!\!\!\!\!\!\!\!\!-\frac{i}{4}{A^0}_\ka\p_\mu\Phi^0{A^0}_\la'{A^0}_\nu'
+\frac{1}{8}{A^0}_\ka\p_\mu\Phi^0{F^0}_{\nu\la}'+\frac{i}{4}{A^0}_\mu
{A^0}_\ka\p_\nu\Phi^0{A^0}_\la'\nn\\&&\!\!\!\!\!\!\!\!\!\!\!\!\!\!\!\!
+\frac{1}{8}{F^0}_{\mu\ka}\p_\nu\Phi^0{A^0}_\la').
 \eea
It should be noted that the Seiberg-Witten maps can not be obtained
uniquely. Indeed the maps for each noncommutative fields are
arbitrary up to a solution of the corresponding homogeneous
equations with undetermined coefficient. However, the physical
results don't depend on this freedom, therefore, those terms which
are solutions of the homogeneous equation are physically irrelevant
\cite{lm}.
\section{Constructing LCNCSM Action}
Lorentz conserving noncommutative standard model can be
constructed in three steps:
\begin{itemize}
 \item Replacing the ordinary products with the star
products.
\item Substituting the noncommutative fields for each
corresponding commutative one.
\item Performing the trace with respect to the
noncommutative tensor with even weight function to make the theory
Lorentz invariant.
\end{itemize}
As is mentioned in the previous section, the noncommutative fields
cannot be uniquely determined by the Seiberg-Witten map i.e. there
is freedom in construction of the noncommutative gauge parameter and
fields by the Seiberg-Witten map. Therefore, the fields can be
appropriately redefined to neglect physically irrelevant terms in
the action. In this paper we extend the electroweak sector of the
standard model to the noncommutative space. One can easily follow
the three steps prescription to derive the action of the LCNCSM for
this sector. To this end we separate the action into four parts as
follows
 \bea
S_{\text{LCNCSM}}=S_{\text{fermion}}+S_{\text{gauge}}+S_{\text{Higgs}}+S_{\text{yukawa}},\label{action}
 \eea
in which each terms will be explained in what follows.
\begin{itemize}
\item $S_{\text{fermion}}$:
\end{itemize}

  This part describes the fermion interaction in the
electroweak sector of the LCNCSM and can be easily written as
 \bea
S_{\text{fermion}}=\int d^6\th\int
d^4xW(\th)(\overline{\widehat{L}}*i\widehat{{\cal
D}}\!\!\!\!/\widehat{L}+\overline{\widehat{R}}*i\widehat{{\cal
D}}\!\!\!\!/\widehat{R})\label{sfermion},
 \eea
  with $\widehat{L}=\widehat{L}_l$ or $\widehat{L}_Q$ where
 \beq
  \widehat{L}_l=\left(
\begin{array}[]{c}
\widehat{\Psi}_{L_{\nu_l}}\\ \widehat{\Psi}_{L_l}
\end{array}
\right)\,\,\,\, , \,\,\,\,\widehat{L}_Q=\left( \begin{array}[]{c}
\widehat{\Psi}_{L_u}\\ \widehat{\Psi}_{L_d}
\end{array} \right) \,\,\, ,
  \label{l}\eeq
and
 \bea
 \widehat{R}=\widehat{\Psi}_{R_l}\,\,\,\, ,
\,\,\,\,\widehat{\Psi}_{R_u}\,\,\,\, , \,\,\,\
\widehat{\Psi}_{R_d},
 \label{r}\eea
 in which subscripts $u$ and $d$, respectively, refer to up-type and down-type
 quarks for all generations and the subscripts $l$ and $Q$ stand for the
 leptons and quarks, respectively.  The covariant derivative $\widehat{{\cal D}}_\mu$
 in terms of the gauge fields $W_\mu$, $B_\mu$ and $G_\mu$ is defined as follows
  \beq \label{cod}
\widehat{{\cal
D}}_\mu\widehat{L}=\left(\p_\mu-igT^a\widehat{W}^a_\mu-ig'\frac{Y}{2}
\widehat{B}_\mu-ig_sT^a_s\widehat{G^a_\mu}\right)\widehat{L},
 \eeq
 and
 \beq \label{codr}
\widehat{{\cal D}}_\mu\widehat{R}=\left(\p_\mu-ig'\frac{Y}{2}
\widehat{B}_\mu-ig_sT^a_s\widehat{G^a_\mu}\right)\widehat{R},
 \eeq
 where $T^a_s$, $T^a$ and $\frac{Y}{2}$ are the generators of
the gauge groups  SU(3)$_C$, SU(2)$_L$ and U(1)$_Y$,
respectively.  The gauge eigenstate weak bosons are related to
the mass eigenstates ( i.e. the electroweak gauge bosons
$(W^\pm,Z)$ and the photon $(A)$ ) by
 \bea
W^\pm_\mu\!\!\!\!&=&\!\!\!\!\frac{W^1_\mu\mp W^2_\mu}{\sqrt{2}},\\
Z_\mu\!\!\!\!&=&\!\!\!\!\frac{-g'B_\mu+gW^3_\mu}{\sqrt{g^2+g'^2}}
=-\sin\th_WB_\mu+\cos\th_WW^3_\mu,\\
A_\mu\!\!\!\!&=&\!\!\!\!\frac{gB_\mu+g'W^3_\mu}{\sqrt{g^2+g'^2}}
=cos\th_WB_\mu+\sin\th_WW^3_\mu,
 \eea
 where the electric charge is $e=g\sin\th_w=g'\cos\th_w$.
 It should be noted that the covariant derivative acts on the fermion or Higgs fields
according to their representations given in Table (1). Furthermore,
the fields with hat are the noncommutative fields and should be
replaced by the appropriate expressions in terms of the ordinary
fields which is obtained up to the second order of $\th_{\mu\nu}$ by
using the Seiberg-Witten maps in Eqs.(\ref{psi1}-\ref{psi22}),
Eqs.(\ref{a1}) and (\ref{a2}) and Eqs.(31-33) for the fermion field,
gauge boson and Higss fields, respectively. One can also see that
the commutative gauge potential $A^0_\mu$ appears in the expansion
of all quantities of noncommutative gauge theory, therefore for the
matter fields we have to use the appropriate vector fields
corresponding to their representations. Namely
\bea\widehat{L}_l[L^0_l,A^0_\mu]\!\!\!\!&=&\!\!\!\!\widehat{L}_l[L^0_l,gT^a
W^a_\mu+g'\frac{Y}{2}B_\mu],\\
\widehat{L}_Q[L^0_Q,A^0_\mu]\!\!\!\!&=&\!\!\!\!\widehat{L}_Q[L^0_Q,g_sT^a_sG^a_\mu+gT^a
W^a_\mu+g'\frac{Y}{2}B_\mu],\\
\widehat{\Psi}_{R_l}[\Psi_{R_l},A^0_\mu]\!\!\!\!&=&\!\!\!\!\widehat{\Psi}_{R_l}[\Psi_{R_l},g'\frac{Y}{2}B_\mu],\\
\widehat{\Psi}_{R_Q}[\Psi_{R_Q},A^0_\mu]\!\!\!\!&=&\!\!\!\!
\widehat{\Psi}_{R_Q}[\Psi_{R_Q},g_sT^a_sG^a_\mu+g'\frac{Y}{2}B_\mu],
 \eea
 where $A^0_\mu$ for the left handed sector is
 \beq
A^0_\mu=\frac{1}{2}g'YB_\mu+gT^aW^a_\mu+g_sT^a_sG^a,
 \eeq
 and for the right handed sector is
 \beq
A^0_\mu=\frac{1}{2}g'YB_\mu+g_sT^a_sG^a.
 \eeq
\begin{table}
\begin{center}
\begin{tabular}{|c|c|c|c|}\hline
    & $SU(3)_c$ & $SU(2)_L$ & $U(1)_Y$ \\ \hline
  $e_R$ & 1 & 1 & -2 \\ \hline
  $L_l=\begin{pmatrix}
    \nu_L \\
    e_L \
  \end{pmatrix}$ & 1 & 2 & -1 \\ \hline
  $u_R$ & 3 & 1 & $\frac{4}{3}$ \\ \hline
  $d_R$ & 3 & 1 & $\frac{-2}{3}$ \\ \hline
  $L_q=\begin{pmatrix}
    u_L \\
    d_L \
  \end{pmatrix}$ & 3 & 2 & $\frac{1}{3}$ \\ \hline
  $\Phi=\begin{pmatrix}
    \phi^+ \\
    \phi^0 \
  \end{pmatrix}$ & 1 & 2 & 1 \\ \hline
\end{tabular}
 \end{center}
 \caption{ Matter and Higgs fields content of the standard model and their representations.}
 \end{table}
 Now inserting the appropriate expansion of the noncommutative fields in terms
 of the ordinary fields in (\ref{sfermion}) leads to the following equation for
  $S_{fermion}$ at the leading order of $\th_{\mu\nu}$ as:
\bea
 \label{fermion}S_{\text{fermion}}=&&\!\!\!\!\!\!\!\!\!\int
d^4x
(\overline{L}i{\cal{D}}\!\!\!\!/\,L+\overline{R}i{\cal{D}}\!\!\!\!/\,R)
+\int d^6\th\int d^4x\th^{\mu\nu}\th^{\ka\la}W(\th)
\Big(-\frac{i}{8}\overline{L}\gamma^\rho
F^0_{\mu\ka}F^0_{\la\rho}{\cal{D}}^0_\nu L\nn\\
&&\!\!\!\!\!\!\!\!\!\! -\frac{i}{4}\overline{L}\gamma^\rho
F^0_{\mu\rho}F^0_{\nu\ka}{\cal{D}}^0_\la L
-\frac{1}{8}\overline{L}\gamma^\rho({\cal{D}}^0_\mu F^0_{\ka\rho}
) {\cal{D}}^0_\nu{\cal{D}}^0_\la L
-\frac{i}{8}\overline{L}\gamma^\rho
 F^0_{\mu\nu}  F^0_{\ka\rho}{\cal{D}}^0_\la L\Big)\nn\\
 &&\!\!\!\!\!\!\!\!\!\!+\int d^6\th\int d^4x\th^{\mu\nu}\th^{\ka\la}W(\th)
\Big(-\frac{i}{8}\overline{R}\gamma^\rho
F^0_{\mu\ka}F^0_{\la\rho}{\cal{D}}^0_\nu R
 -\frac{i}{4}\overline{R}\gamma^\rho
F^0_{\mu\rho}F^0_{\nu\ka}{\cal{D}}^0_\la R\nn\\
&&\!\!\!\!\!\!\!\!\!\!-\frac{1}{8}\overline{R}\gamma^\rho({\cal{D}}^0_\mu
F^0_{\ka\rho} ) {\cal{D}}^0_\nu{\cal{D}}^0_\la R
-\frac{i}{8}\overline{R}\gamma^\rho
 F^0_{\mu\nu}  F^0_{\ka\rho}{\cal{D}}^0_\la R\Big).
 \eea
In obtaining (\ref{fermion}) the irrelevant terms is ignored by
redefinition of the fields via the freedom in determining of the
Seiberg-Witten maps.  Equation (\ref{fermion}) shows that besides
the usual Standard Model and the NCSM interactions, there are new
couplings between the fermions and the electroweak gauge bosons such
as $ff\gamma\gamma\gamma$, $ff\gamma\gamma Z$, $ff\gamma ZZ$ and
$ffZZZ$ and so on.  The vertex $ff\gamma\gamma\gamma$ is one of the
vertices of LCNCQED \cite{ccz,he}. In the noncommutative space, a
neutral particle can interact with photon in the adjoint
representation \cite{hez,np}. These interactions are proportional to
the odd power of $\th_{\mu\nu}$, therefore they are absent in the
Lorentz invariant noncommutative field theory, see the action
(\ref{fermion}). Nevertheless in contrast to the minimal NCSM there
is not photon-neutrino coupling in the LCNCSM \cite{hez}.
 \begin{itemize}
\item $S_{\text{gauge}}$:
\end{itemize}
 This term contains the kinetic terms for the gauge bosons
 of the standard model. The general form of the gauge invariant
action for the gauge sector of the LCNCSM can be written as follows
 \beq
 S_{\text{gauge}}=-\frac{1}{2}\int d^6\th\int d^4x W(\th)
Tr\{\frac{1}{G^2}\widehat{F}^{\mu\nu}*\widehat{F}_{\mu\nu}\},\label{tr}
 \eeq
where $Tr$ is trace over all representations.  $G$ is an operator
which determines the coupling constants of the theory and
commutes with all generators of $SU(2)$ and $SU(3)$ and is
defined as \cite{sm}
 \beq
 {\frac{1}{g^2_I}}=Tr\{\frac{1}{G^2}T^a_IT^a_I\},
 \eeq
 where $g_I$ and $T^a_I$ are the ordinary coupling constants and generators
 of the gauge group, respectively.
Since the gauge group is extended to incorporate the noncommutative
corrections, according to the Seiberg-Witten map and using the
enveloping algebra \cite{ncg}, one encounters all ordered tensor
powers of the generators $T^a$ in the trace in (\ref{tr}).
Therefore, the trace in the kinetic terms for gauge bosons in
contrast to the ordinary case is not unique and depends on a choice
of the representation of the gauge group \cite{sm,gut,ztog}. The
minimal choice can be the simplest choice in which $Tr$ is a sum of
three traces over the $U(1)$, $SU(2)$ and $SU(3)$ sectors with the
fundamental representations for $SU(2)$ and $SU(3)$ generators in
the corresponding traces and \cite{sm}
 \bea
Y&=&\begin{pmatrix}
  1 & 0 \\
  0 &-1
\end{pmatrix}.
 \eea
However, the freedom in the choice of the traces can be used to
construct new versions of the LCNCSM. Since the fermion-gauge boson
interactions remain the same regardless on the choice of traces in
the gauge sector, the matter sector of the action is the same for
all versions of the LCNCSM.  Nevertheless, in the non-minimal
versions of the theory new parameters appear which cannot be
uniquely obtained in the theory
\cite{sm,ztog}.\\
The general form of $S_{\text{gauge}}$ in terms of SM fields can be
obtained by inserting the expansion of the field strength
$\widehat{F}_{\mu\nu}=\p_\mu\widehat{A}_\nu-\p_\nu\widehat{A}_\mu-i
[\widehat{A}_\mu,\widehat{A}_\nu]$ in terms of the commutative one
in (\ref{tr}) as follows
 \bea\label{gaugeaction}
S_{\text{gauge}}=-\frac{1}{2}\int d^6\th\int d^4x
&&\!\!\!\!\!\!\!\!\!\!W(\th)Tr\{\frac{1}{G^2}\{-\frac{1}{2}F^0_{\mu\nu}{F^0}^{\mu\nu}\nn\\&&\!\!\!\!\!\!\!\!\!\!\!\!\!\!\!\!\!\!\!\!\!\!\!\!\!\!\!\!\!\!
+\th^{\mu\nu}\th^{\ka\la}\big(\frac{1}{8}
 F^0_{\mu\nu}F^0_{\ka\la}F^0_{\rho\si} F^{0 \rho\si}-\frac{i}{4}
 F^0_{\mu\nu}(\mathcal{D}^0_\ka F^0_{\rho\si})(\mathcal{D}^0_\la  F^{0
 \rho\si})\nn\\&&\!\!\!\!\!\!\!\!\!\!- \frac{1}{8}(\mathcal{D}^0_\mu\mathcal{D}^0_\ka F^0_{\rho\si})
 (\mathcal{D}^0_\nu\mathcal{D}^0_\la F^{0 \rho\si})+
 \frac{i}{2}(\mathcal{D}^0_\mu F^0_{\rho\ka})(\mathcal{D}^0_\nu F^0_{\si\la})F^{0 \rho\si}\nn\\&&\!\!\!\!\!\!\!\!\!\!+\frac{1}{2}
 F^0_{\mu\rho}F^0_{\nu\si}F^{0\hspace{1mm}\rho}_{\ka}F^{0\hspace{1mm}\si}_{\la}+\frac{1}{2}
 F^0_{\mu\rho}F^0_{\nu\si}F^{0\hspace{1mm}\si}_{\ka}F^{0\hspace{1mm}\rho}_{\la}\nn\\&&\!\!\!\!\!\!\!\!\!\!-\frac{1}{2}
 F^0_{\mu\nu}F^0_{\ka\rho}F^0_{\la\si} F^{0 \rho\si}-\frac{1}{2}
F^0_{\ka\rho}F^0_{\la\si} F^0_{\mu\nu} F^{0
\rho\si}\\&&\!\!\!\!\!\!\!\!\!\!+\frac{1}{2}(F^0_{\mu\ka}F^0_{\nu\rho}F^0_{\la\si}+2F^0_{\nu\rho}F^0_{\mu\ka}F^0_{\la\si}+F^0_{\la\si}F^0_{\nu\rho}F^0_{\mu\ka})F^{0
\rho\si}\big)\}\},\nn
 \eea
 where $F^0_{\mu\nu}=\p_\mu A^0_\nu-\p_\nu
A^0_\mu-i[A^0_\mu,A^0_\nu]$ is the field strength in the usual
space and
 \beq
  A_\mu=\frac{1}{2}g'Y B_\mu
+gT^aW^a_\mu+g_sT^a_s{G^a_s}_\mu\label{amu}.
 \eeq
 In equation (\ref{gaugeaction}) there are new vertices in all versions
 of LCNCSM in comparison  with the commutative standard model. Meanwhile, in contrast to the minimal noncommutative
  standard model, new interactions appear in the electroweak part of the LCNCSM at the leading order of $\th_{\mu\nu}$.
  However, inserting the electroweak part of (\ref{amu}) in (\ref{gaugeaction}) and performing the trace operation
 in the minimal case lead to the electroweak sector of the minimal version of $S_{gauge}$:

 \bea
  S_{gauge}^{mLCNCSM}&=&-\frac{1}{2}\int d^6\th\int d^4x
W(\th)\Big(\frac{1}{2}B_{\mu\nu}B^{\mu\nu}+W^a_{\mu\nu}{W^a}^{\mu\nu}\nn\\&+&
\th^{\mu\nu}\th^{\ka\la}\Big({g^\prime}^2\Big(\frac{1}{64}
 B_{\mu\nu}B_{\ka\la}B_{\rho\si}-\frac{1}{8}
 B_{\mu\nu}B_{\ka\rho}B_{\la\si}+\frac{1}{4}
 B_{\mu\ka}B_{\nu\rho}B_{\la\si}\Big)B^{\rho\si}\nn\\&+ &
 \frac{g^2}{4}\Big(\frac{1}{8}
 ({W^a}_{\mu\nu}{W^a}_{\ka\la})({W^b}_{\rho\si}{W^b}^{\rho\si})
 +({W^a}_{\mu\rho}{{W^a}_\la}^{\rho})({W^b}_{\nu\si}{{W^b}_\ka}^{\si})\nn\\
 &+&({W^a}_{\mu\rho}{{W^a}_\la}^{\si})({W^b}_{\nu\si}{{W^b}_\ka}^{\rho})
 -({W^a}_{\mu\nu}{W^a}^{\rho\si})({W^b}_{\ka\rho}{W^b}_{\la\si})\nn\\
 &+&2({W^a}_{\mu\ka}{W^a}_{\nu\rho})({W^b}_{\la\si}{W^b}^{\rho\si})\Big)+g^\prime
 g\Big(\frac{1}{32}(B_{\mu\nu}B_{\ka\la})(W^a_{\rho\sigma}{W^a}^{\rho\sigma})\nn\\
 &+&\frac{1}{32}(B_{\rho\sigma}B^{\rho\sigma})(W^a_{\mu\nu}{W^a}_{\ka\la})
 +\frac{1}{8}(B_{\mu\nu}B_{\rho\si})(W^a_{\ka\la}{W^a}^{\rho\si})\nn\\
 &-&\frac{1}{2}(B_{\mu\nu}B_{\ka\rho})(W^a_{\la\si}{W^a}^{\rho\si})
 -\frac{1}{2}(B_{\la\si}B^{\rho\si})(W^a_{\mu\nu}{W^a}_{\ka\rho})\nn\\
 &-&\frac{1}{4}(B_{\mu\nu}B^{\rho\si})(W^a_{\ka\rho}{W^a}_{\la\si})
 -\frac{1}{4}(B_{\ka\rho}B_{\la\si})(W^a_{\mu\nu}{W^a}^{\rho\si})\nn\\
 &+&\frac{1}{2}(B_{\mu\ka}B^{\rho\si})(W^a_{\nu\rho}{W^a}_{\la\si})
 +\frac{1}{2}(B_{\nu\rho}B_{\la\si})(W^a_{\mu\ka}{W^a}^{\rho\si})\nn\\
 &+&(B_{\mu\ka}B_{\nu\rho})(W^a_{\la\si}{W^a}^{\rho\si})
 +(B_{\la\si}B^{\rho\si})(W^a_{\mu\ka}{W^a}_{\nu\rho})\Big)\nn\\
 &-&\frac{g^3}{8}\ep^{abc}W^a_\nu\p_\la{W^b}^{\rho\si}\p_\mu\p_\ka W^c_{\rho\si}
-\frac{g^4}{16}\ep^{abc}\ep^{dce}W^a_\ka{W^b}^{\rho\si}W^d_\nu\p_\mu\p_\ka
W^e_{\rho\si}\nn\\
&-&\frac{g^4}{16}\ep^{abc}\ep^{dec}\p_\mu(W^a_\ka{W^b}^{\rho\si})W^d_\nu\p_\la
W^e_{\rho\si}
-\frac{g^4}{16}\ep^{abc}\ep^{dec}W^a_\mu\p_\ka{W^b}^{\rho\si})\p_\nu(W^d_\ka
W^e_{\rho\si}\nn\\
&-&\frac{g^4}{16}\ep^{abc}\ep^{dec}W^a_\mu\p_\ka{W^b}^{\rho\si})W^d_\nu\p_\la
W^e_{\rho\si}
-\frac{g^4}{16}\ep^{abc}\ep^{dec}W^a_\ka{W^b}^{\rho\si})W^d_\mu\p_nu\p_\la
W^e_{\rho\si}\nn\\
&-&\frac{g^5}{16}\ep^{abc}\ep^{dce}\ep^{fge}W^a_\la{W^b}^{\rho\si}W^d_\nu
W^f_\mu \p_\ka W^g_{\rho\si}
-\frac{g^5}{16}\ep^{abc}\ep^{dce}\ep^{fge}W^a_\la{W^b}^{\rho\si}W^d_\nu
\p_\mu(W^f_\ka W^g_{\rho\si})\nn\\
&-&\frac{g^5}{16}\ep^{abc}\ep^{dce}\ep^{fge}W^a_\ka{W^b}^{\rho\si}W^d_\mu\p_\nu(
W^f_\la W^g_{\rho\si})
-\frac{g^5}{16}\ep^{abc}\ep^{dce}\ep^{fge}W^a_\ka{W^b}^{\rho\si}W^d_\mu
W^f_\nu \p_\la W^g_{\rho\si}\nn\\
&-&\frac{g^6}{16}\ep^{abc}\ep^{dce}\ep^{fgh}\ep^{ihe}W^a_\ka{W^b}^{\rho\si}W^d_\mu
W^f_\ka W^g_{\rho\si}W^i_\nu
 \Big)\label{minigaugeaction},\eea
where for the hypercharge,  $B_{\mu\nu}$ is defined as
 \beq
B_{\mu\nu}=\p_\mu B_\nu-\p_\nu B_\mu,
 \eeq
and for SU(2) gauge fileds we define
 \beq
W^a_{\mu\nu}=\p_\mu W^a_\nu-\p_\nu W^a_\mu+g\ep^{abc}W^b_\mu
W^c_\nu.
 \eeq
 As (\ref{minigaugeaction}) shows in the minimal version of LCNCSM
 (mLCNCSM) at the lowest order there are vertices such as
 $\gamma\gamma\gamma\gamma$, $\gamma\gamma\gamma Z$, $\gamma\gamma Z
 Z$, $\gamma Z Z Z$, $ZZZZ$ and so on.
\begin{itemize}
\item $S_{\text{Higgs}}$:
\end{itemize}
 This part of the action is responsible for breaking of the $SU(2)_L\times U(1)_Y$ gauge
 symmetry of the standard
model via the Higgs mechanism which in turn generates masses for the
gauge bosons. The Higgs action in the Lorentz invariant
noncommutative space can be written as follows
  \beq
S_{\text{Higgs}}=\int d^6\th\int d^4x
W(\th)\big((\widehat{\mathcal{D}}_\mu\widehat{\Phi})^\dagger*(\widehat{\mathcal{D}}^\mu\widehat{\Phi})
-\mu^2\widehat{\Phi}^\dagger*\widehat{\Phi}
-\la(\widehat{\Phi}^\dagger*\widehat{\Phi})^2\big),
 \eeq
 where the non-commutative Higgs field is given by the hybrid
Seiberg-Witten map which is already obtained up to the second order
of $\th_{\mu\nu}$ in the previous section. In the leading order of
the expansion in $\th_{\mu\nu}$ one explicitly obtains
 \bea
 S_{\text{Higgs}}=\int d^6\th\int d^4x
W(\th)\Big(&&\hspace{-7mm}({\cal{D}}^0_\al\Phi^0)^\dag({{\cal{D}}^0}^\al\Phi^0)
-\mu^2{\Phi^0}^\dag\Phi^0-\la({\Phi^0}^\dag\Phi^0)^2
\nn\\
&&\hspace{-25mm}+\th^{\mu\nu}\th^{\ka\la}\Big(({\cal{D}}^0_\al\Phi^0)^\dag({\cal{D}}^0_\al\Phi^2_{\mu\nu\ka\la})
+({\cal{D}}^0_\al\Phi^2_{\mu\nu\ka\la})^\dag({\cal{D}}^0_\al\Phi^0)
+({\cal{D}}^0_\al\Phi^1_{\mu\nu})^\dag({\cal{D}}^0_\al\Phi^1_{\ka\la})\nn\\
&&\hspace{-25mm} -i({\cal{D}}^0_\al\Phi^0)^\dag
{{A^2}^\al}_{\mu\nu\ka\la}\Phi^0+i(\Phi^0)^\dag
{{A^2}^\al}_{\mu\nu\ka\la}({\cal{D}}^0_\al\Phi^0)-i({\cal{D}}^0_\al\Phi^0)^\dag
{{A^1}^\al}_{\mu\nu}\Phi^1_{\ka\la}\nn\\
&&\hspace{-25mm} +i(\Phi^1_{\mu\nu})^\dag
{{A^1}^\al}_{\ka\la}({\cal{D}}^0_\al\Phi^0)+i{\Phi^0}^\dag
{{A^1}^\al}_{\mu\nu}({\cal{D}}^0_\al\Phi^1_{\ka\la})
-i({\cal{D}}^0_\al\Phi^1_{\mu\nu})^\dag {{A^1}^\al}_{\ka\la}\Phi^0\nn\\
&&\hspace{-25mm} +{\Phi^0}^\dag
{{A^1}^\al}_{\mu\nu}{{A^1}_\al}_{\ka\la}\Phi^0\Big)\Big)\label{higgsaction}.
 \eea
Here $\mathcal{D}_\mu$ can be appropriately defined very similar to
the (\ref{cod}) and (\ref{codr}) according to the representations
given in Table (1).  The functions $\Phi^1_{\mu\nu} $,
$\Phi^2_{\mu\nu\ka\la}$, ${A^1_\al}_{\mu\nu}$ and
${A^2_\al}_{\mu\nu\ka\la}$ can be easily obtained by comparing
(\ref{phi1}), (\ref{phi2}), (\ref{a1}) and (\ref{a2}), respectively,
with
 \bea
\Phi^1(\Phi^0,A^0_\mu,&&\hspace{-7mm}{A^0}^\prime_\mu)=\th^{\mu\nu}\Phi^1_{\mu\nu}(\Phi^0,\frac{g^\prime}{2}YB_\mu+gT^aW^a_\mu,0)\,\,\,,\nn\\
\Phi^2(\Phi^0,A^0_\mu,&&\hspace{-7mm}{A^0}^\prime_\mu)=\th^{\mu\nu}\th^{\ka\la}\Phi^2_{\mu\nu\ka\la}
(\Phi^0,\frac{g^\prime}{2}YB_\mu+gT^aW^a_\mu,0)\,\,\,,\nn\\
A^1_\al(A^0_\mu)=&&\hspace{-5mm}\th^{\mu\nu}{A^1_\al}_{\mu\nu}(\frac{g^\prime}{2}YB_\mu+gT^aW^a_\mu)\,\,\,,\nn\\
A^2_\al(A^0_\mu)=&&\hspace{-5mm}\th^{\mu\nu}\th^{\ka\la}{A^2_\al}_{\mu\nu\ka\la}(\frac{g^\prime}{2}YB_\mu+gT^aW^a_\mu).
 \eea
The $\th_{\mu\nu}$-independent part of (\ref{higgsaction}) is the
same as
  the usual action for the Higgs part of the standard model.
  Meanwhile, the remaining part of the action contains the
  derivative of the Higgs field (which does not have contribution to the minimum value of
the potential) and terms containing the gauge and Higgs fields both.
The latter terms are also vanished by fixing the vacuum expectation
value.  Therefore, the spontaneous symmetry breaking occurs
according to the commutative standard model but in contrast to the
standard model, numerous new couplings between the Higgs and the
electroweak gauge bosons appear in this theory.  One can easily see
from (\ref{higgsaction}) that among these
 new interactions  there are couplings solely between the gauge fields with the coupling strength
proportional to $(\langle\Phi\rangle_0)^2$.
\begin{itemize}
\item $S_{\text{yukawa}}$:
\end{itemize}
 This part describes  the Yukawa interactions between fermions and Higgs field which lead to
the mass generation for fermions after the symmetry breaking. The
Yukawa action of LCNCSM can be written as follows
 \beq
 S_{\text{yukawa}}=-\int d^6\th\int d^4x
W(\th)\big(G_{ij}\overline{\widehat{L}}^i*\widehat{\Phi}*\widehat{R}^j
+G_{ij}\overline{\widehat{R}}^i*\widehat{\Phi}^\dag*\widehat{L}^j\big),
 \eeq
where $i$ and $j$ refer to the different generations. Once again the
noncommutative fields have to be expanded in terms of the
corresponding ordinary fields up to the second order of
$\th_{\mu\nu}$ but one should note that $\widehat{\Phi}$ has to be
appropriately written with respect to the representation of its left
and right fields according to Table (1). After some algebra, the
general form for the Yukawa action, up to the leading order is
 \bea S_{\text{yukawa}}=-\int
d^6\th\int d^4x&&\hspace{-7mm}
W(\th)G_{ij}\Big(\overline{L}^i\Phi{R}^j+\th^{\mu\nu}\th^{\ka\la}\Big(\Lo{\Phi}^1_{\mu\nu}{R^1}^j_{\ka\la}
+\Loo_{\mu\nu}\Phi^1_{\ka\la}R^j\nn\\
&&\hspace{-5mm}+\Loo_{\mu\nu}\Phi{R^1}^j_{\ka\la}+\Lot_{\mu\nu\ka\la}\Phi
R^j+\Lo\Phi{{R^2}^j}_{\mu\nu\ka\la}+\Lo\Phi^2_{\mu\nu\ka\la}R^j\nn\\
&&\hspace{-5mm}+\frac{i}{2}\Loo_{\mu\nu}\p_\ka\Phi\p_\la
R^j+\frac{i}{2}\Lo\p_\mu\Phi^1_{\ka\la}\p_\nu
R^j+\frac{i}{2}\Lo\p_\mu\Phi\p_\nu
{{R^1}^j}_{\ka\la}\nn\\
&&\hspace{-5mm}-\frac{1}{8}\Lo\p_\mu\p_\ka\Phi\p_\nu\p_\la
R^j\Big)\Big)+\text{C. C.},
 \label{yukaction}\eea
 where $L$, $R$ and $\Phi$ are ordinary fields and for leptons we
 define
  \bea
 R^1(R^0,A^0_\mu)=&&\hspace{-5mm}\th^{\mu\nu}R^1_{\mu\nu}(R^0,\frac{g^\prime}{2}YB_\mu),\nn\\
 R^2(R^0,A^0_\mu)=&&\hspace{-5mm}\th^{\mu\nu}\th^{\ka\la}R^2_{\mu\nu\ka\la}(R^0,\frac{g^\prime}{2}YB_\mu),\nn\\
 L^1(L^o,A^0_\mu)=&&\hspace{-5mm}\th^{\mu\nu}L^1_{\mu\nu}(L^0,\frac{g^\prime}{2}YB_\mu+gT^aW^a_\mu),\nn\\
 L^2(L^o,A^0_\mu)=&&\hspace{-5mm}\th^{\mu\nu}\th_{\ka\la}L^2_{\mu\nu\ka\la}(L^0,\frac{g^\prime}{2}YB_\mu+gT^aW^a_\mu),\nn\\
\Phi^1(\Phi^0,A^0_\mu,A'^0_\mu)=&&\hspace{-5mm}\th^{\mu\nu}\Phi^1_{\mu\nu}(\Phi^0,\frac{1}{2}g'YB_\mu
+gT^aW^a_\mu,\frac{1}{2}g'YB_\mu),\nn\\
\Phi^2(\Phi^0,A^0_\mu,A'^0_\mu)=&&\hspace{-5mm}\th^{\mu\nu}\th^{\ka\la}\Phi^2_{\mu\nu\ka\la}(\Phi^0,\frac{1}{2}g'YB_\mu
+gT^aW^a_\mu,\frac{1}{2}g'YB_\mu),\label{yuklepton}
 \eea
 while for quarks  $g_sT^a_sG^a_s$ should be added to gauge
 fields to define
 \bea\label{yukquark}
 R^1(R^0,A^0_\mu)=&&\hspace{-5mm}\th^{\mu\nu}R^1_{\mu\nu}(R^0,\frac{g^\prime}{2}YB_\mu+g_sT^a_sG^a_\mu),\nn\\
 R^2(R^0,A^0_\mu)=&&\hspace{-5mm}\th^{\mu\nu}\th^{\ka\la}R^2_{\mu\nu\ka\la}(R^0,\frac{g^\prime}{2}YB_\mu+g_sT^a_sG^a_\mu),\nn\\
 L^1(L^o,A^0_\mu)=&&\hspace{-5mm}\th^{\mu\nu}L^1_{\mu\nu}(L^0,\frac{g^\prime}{2}YB_\mu+gT^aW^a_\mu+g_sT^a_sG^a_\mu),\nn\\
 L^2(L^o,A^0_\mu)=&&\hspace{-5mm}\th^{\mu\nu}\th{\ka\la}L^2_{\mu\nu\ka\la}(L^0,\frac{g^\prime}{2}YB_\mu+gT^aW^a_\mu+g_sT^a_sG^a_\mu),\nn\\
\Phi^1(\Phi^0,A^0_\mu,A'^0_\mu)=&&\hspace{-5mm}\th^{\mu\nu}\Phi^1_{\mu\nu}(\Phi^0,\frac{1}{2}g'YB_\mu
+gT^aW^a_\mu+g_sT^a_sG^a,\frac{1}{2}g'YB_\mu+g_sT^a_sG^a),\nn\\
\Phi^2(\Phi^0,A^0_\mu,A'^0_\mu)=&&\hspace{-5mm}\th^{\mu\nu}\th^{\ka\la}\Phi^2_{\mu\nu\ka\la}(\Phi^0,\frac{1}{2}g'YB_\mu
+gT^aW^a_\mu+g_sT^a_sG^a,\frac{1}{2}g'YB_\mu+g_sT^a_sG^a).\nn\\
 \eea
 $L$ and $R$ in the left hand sides of (\ref{yuklepton}) and (\ref{yukquark}) are defined in (\ref{l}) and (\ref{r})
 and can be used to obtain $L^1_{\mu\nu}$, $R^1_{\mu\nu}$ and so on by expanding the left hand side of the equations up to the
 desired order of $\th_{\mu\nu}$ and comparing the result with the right hand sides of (\ref{yuklepton}) and (\ref{yukquark}).
  One can easily see that (\ref{yukaction}) reproduces its counterpart in the standard model
   and many new couplings between the standard model fields.
   Finally, we note that in the LCNCSM, as in the Standard Model, a Cabibbo Kobayashi Maskawa matrix
in the charged currents can be obtained by diagonalizing the Yukawa
coupling matrices using biunitary transformations.

\section{The Phenomenological Test of LCNCSM }
In constructing LCNCSM one finds besides the usual standard model,
many new couplings between the ordinary fields of the standard
model. For instance, here one has four gauge boson couplings such as
4-$\gamma$, 4-$Z$, ...; fermion gauge boson couplings such as
$ff\gamma\gamma$, $ff\gamma\gamma\gamma$, ... and so on.
Furthermore, each usual vertex in the standard model receives
corrections from the LCNCSM. Therefore, there are many measurable
quantities in the LCNCSM which can show, if exist, the effects of
noncommutative space in future experiments. To this end we consider
the neutral current interaction with $Z_0$ for leptons which can
read from (\ref{fermion}) as follows
 \bea
 \int d^6\th\int
d^4xW(\th)\frac{g}{2cos\th}\Big(&&\!\!\!\!\!\!\!\!\!
\overline{\nu}_LZ_0\!\!\!\!\!\!/\,\,\,\nu_L-
\frac{1}{8}\th^{\mu\nu}\th^{\ka\la}\overline{\nu}_L\ga^\rho\p_\mu
{Z_0}_{\ka\rho}\p_\nu\p_\la\nu_L
+(2sin^2\th-\frac{1}{2})\overline{e}Z_0\!\!\!\!\!\!/\,\,\,
e\nn\\&&\!\!\!\!\!\!\!\!\!\!\!\!\!\!\!\!\!\!
-\frac{1}{8}\th^{\mu\nu}\th^{\ka\la}(2sin^2\th-\frac{1}{2})\overline{e}\ga^\rho\p_\mu
{Z_0}_{\ka\rho}\p_\nu\p_\la
e+(2sin^2\th-\frac{1}{2})\overline{e}Z_0\!\!\!\!\!\!/\,\,\,\ga_5
e\nn\\&&\!\!\!\!\!\!\!\!\!\!\!\!\!\!\!\!\!\!
-\frac{1}{8}\th^{\mu\nu}\th^{\ka\la}(2sin^2\th-\frac{1}{2})\overline{e}\ga^\rho\ga_5\p_\mu
{Z_0}_{\ka\rho}\p_\nu\p_\la e\Big),
 \eea
  where ${Z_0}_{\mu\nu}=\p_\mu{Z_0}_\nu-\p_\nu{Z_0}_\mu$. This action
leads to the following Feynman rule for $eeZ$-vertex, Fig.(1):

\begin{figure}\centerline{\epsfysize=1.5
in\epsfxsize=1.5 in\epsffile{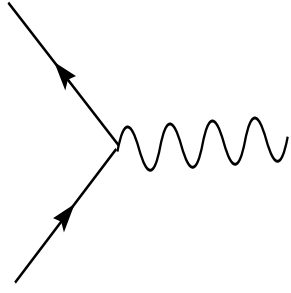}} \caption{
 Zee-vertex.  The bold and wavy lines show the electron and Z gauge boson, respectively.  In the LCNCSM the vertex is corrected with
 respect to the standard model as $\Gamma^{NC}_\mu=\Gamma_\mu\Big(1+\frac{\langle\th^2\rangle}{96}(\frac{k^4}{4}-m^2_f
k^2)\Big)$, see (\ref{fr2}). }
\end{figure}

\beq \label{fr2}
 \frac{i
g}{\sin(2\th_W)}\ga^{\mu}\Big(-2Q_f\sin^2\th_W+
T_3(1-\ga_5)\Big)\Big(1+\frac{\langle\th^2\rangle}{96}(\frac{k^4}{4}-m^2_f
k^2)\Big).
 \eeq
 Thus the vector coupling constant $C_V$ and the scalar one $C_A$ are
 corrected by the new term proportional to $Q^4$. On the other hand according to the implemented
experiments, such dependency was not confirmed. The noncommutative
correction of the coupling constants cannot absorb through their
redefinitions, therefore this correction must be smaller than the
resolution of the experiments. Comparing with \cite{ex}, we find the
noncommutative scale $\La_C=\sqrt{\frac{12}{\langle\th^2\rangle}}$
must be larger than 112 GeV. To find better bound on the
noncommutative scale, we should use linear colliders with a C.M.E.
around few TeV's. The QED sector of the LCNCSM has been investigated
by \cite{he} which leads to $\La_C\sim300 \,\, GeV$ and by \cite{kn}
which leads to $\La_C\sim160 \,\, GeV$.
\section{Conclusions}
 In this paper, we have constructed the Lorentz conserving version
of the noncommutative standard model. For this purpose the
$\th_{\mu\nu}$-expansion of the standard model fields, up to the
second order of $\th_{\mu\nu}$ is obtained as given in
(\ref{psi22}), (\ref{a2}) and (\ref{phi2}).  The Seiberg-Witten map
for the Higgs field, up to the second order of $\th_{\mu\nu}$, is
calculated for the first time, see (\ref{phi2}).  While,
(\ref{psi22}) and (\ref{a2}) have already been calculated  but there
are differences in signs of (\ref{a2}) in comparison with the
corresponding equation given in reference \cite{lm}. The misprinting
can be easily understood by considering the reduction of the
equation to the
 Abelian case, though to find its correct form we have recalculated the equation.
Consequently, the action of the LCNCSM is introduced in terms of
four terms, see (\ref{action}).  It is shown that in all versions of
the LCNCSM new vertices appear in comparison with the ordinary
standard model.  In the minimal version of LCNCSM there are also new
point interactions in contrast to the minimal NCSM.  For instance,
in the minimal LCNCSM  besides the usual Standard Model and the NCSM
interactions, there are new couplings between the fermions and the
electroweak gauge bosons such as $ff\gamma\gamma\gamma$,
$ff\gamma\gamma Z$, $ff\gamma ZZ$ and $ffZZZ$ and so on.  The vertex
$ff\gamma\gamma\gamma$ is one of the vertices of LCNCQED
\cite{ccz,he}.  Nevertheless in contrast to the minimal NCSM there
is not any photon-neutrino coupling in the LCNCSM \cite{hez}. Indeed
such interactions are proportional to the odd power of
$\th_{\mu\nu}$ which
 are absent in the Lorentz invariant noncommutative field theory,
see the action (\ref{fermion}).  Furthermore, It is shown that in
the mLCNCSM at the lowest order there are vertices such as
 $\gamma\gamma\gamma\gamma$, $\gamma\gamma\gamma Z$, $\gamma\gamma Z
 Z$, $\gamma Z Z Z$, $ZZZZ$ and so on, see (\ref{minigaugeaction}).

In constructing LCNCSM  besides many new couplings between the
ordinary fields of the standard model, each usual vertex in the
standard model receives corrections from the LCNCSM. Therefore,
there are many measurable quantities in the LCNCSM which can show
the effects of noncommutative space in future experiments.  However,
among the measurable quantities, the coupling constant $C_V$ and
$C_A$ can be corrected by considering the $eeZ$-vertex which leads
to the value $\La_C\sim 112 GeV$ for the noncommutative scale.

\section*{Acknowledgements}

The financial support of Isfahan University of Technology research
council is acknowledged.


\end{document}